# Azimuthal eigenmodes at strongly non-degenerate parametric down-conversion

L.S. Dvernik, P.A. Prudkovskii

Lomonosov Moscow State University, Moscow, 119991 Russia
(Pavel A. Prudkovskii vysogota@gmail.com)

Quantum-optical technologies based on parametric light down-conversion are not yet applied in the terahertz frequency range. This is owing to the complexity of detecting single photons in the terahertz frequency range and the strong entanglement of modes of optical-terahertz biphotons. This study investigates the angular structure of scattered radiation generated by strongly non-degenerate parametric down-conversion when the frequency of the idler radiation does not exceed several terahertz. We demonstrate that under certain approximations, it is possible to obtain azimuthal eigenmodes for the nonlinear-interaction operator. The solution of the evolution equations for the field operators in these eigenmodes has the form of the Bogolyubov transformation, which allows a scattering matrix to be obtained for arbitrary values of the parametric gain. This scattering matrix describes both the production of biphoton pairs and the generation of intense fluxes of correlated optical-terahertz fields that form a macroscopic quantum state of radiation in two spectral ranges.

## I. INTRODUCTION

The properties of parametric down-conversion (PDC) have been actively studied since the late 1960s. Biphoton states generated by PDC are used in quantum technologies [1], and the relationship between the spectral-angular characteristics of scattered radiation and the dispersion properties of a nonlinear crystal is the basis of spectroscopic methods. The frequencies of a pair of conjugate photons, produced by PDC, can lie in different ranges of the electromagnetic radiation spectrum. This enabled the development of methods to study the dispersion properties of crystals in the infrared region without detecting a field in this frequency range [2-5] and to apply low-energy correlated PDC radiation fluxes in X-ray diagnostics [6].

Over the past 10 years, the possibilities of generating and detecting terahertz radiation using strongly non-degenerate PDC have been actively studied [7]. This has already enabled the development of both methods for terahertz spectroscopy [8-12] and methods for measuring the brightness of terahertz radiation [13-15].

Optical-terahertz quantum states of the electromagnetic field may enable the application of quantum information technologies in the terahertz frequency range. Unfortunately, the quantum correlation between optical and terahertz photons has not yet been reported. All applications realized to date are based on theoretical assumptions and the detection of the optical part of spontaneous PDC (SPDC) radiation. This is primarily owing to the complexity of detecting weak radiation fluxes in the terahertz frequency range, in particular, with the absence of single-photon terahertz detectors. Therefore, to date, only theoretical studies on the properties of optical-terahertz biphotons [16-18] and indirect studies on the quantum properties of the terahertz field generated by PDC have been conducted [19].

Another limitation on the study of optical-terahertz biphoton states is the large wavelength of terahertz radiation. If the wavelength of scattered radiation is of the same order of magnitude as the scattering volume, the optical-terahertz field is in a complex entangled state with a large number of angular modes. In the optical range, in the case of SPDC, mode separation can be conducted using Schmidt decomposition [12,20-23]. However, the large wavelength of terahertz scattered radiation results in a more complex angular shape of the modes. In addition, the absence of appropriate single-photon detectors requires the use of stimulated PDC when the parametric gain is significantly greater than unity. In the general case of an arbitrary gain, the Bloch-Messiah decomposition can be used to separate modes [24]; however, this procedure is more complicated and typically cannot be performed analytically [25].

This study demonstrates that under particular approximations in the case of strongly non-degenerate PDC, the operator of the nonlinear parametric interaction can be diagonalized in the azimuthal angle space, and the azimuthal eigenmodes can be determined. In the basis of azimuthal eigenmodes, it is easy to obtain the scattering matrix for the operators of the scattered radiation field at an arbitrary parametric gain. This scattering matrix enables to describe any quantum properties of scattered radiation, as well as to determine the optimal detection conditions depending on the effective number and shape of azimuthal modes at different terahertz radiation frequencies. Since in our approximations the azimuthal eigenmodes do not depend on the parametric gain, this makes it possible in experiments to smoothly pass from individual biphotons to intense optical-terahertz twin-beams without changing the mode structure of the scattered radiation.

The paper is organized as follows. In Sec. II we consider the complete structure of the nonlinear interaction operator for strongly non-degenerate PDC. This operator describes the evolution of the operators of scattered radiation propagating in a nonlinear crystal. In Sec. III we consider all angular dependences affecting the structure of the scattered radiation modes and investigate the conditions under which the nonlinear interaction operator can be diagonalized in the azimuthal angle space. In Sec. IV we diagonalize the nonlinear interaction operator and obtain azimuthal eigenmodes. This makes it possible to obtain a scattering matrix describing the scattered radiation at an arbitrary parametric gain. Sec. V studies the shape of the azimuthal eigenmodes and shows how the effective number of modes depends on the frequency of terahertz radiation and the parametric gain. Final conclusions are given in Sec. VI.

## II. THE OPERATOR OF THE NONLINEAR INTERACTION

To describe the evolution of the operators of the scattered field in a nonlinear crystal, it is convenient to use the approach presented in [26]. In this approach, the periodicity of the quantized field in the direction of pump propagation is replaced by a periodicity in time, with a period $T$, that significantly exceeds all other time scales of the system. In this case, the negative frequency part of the scattered fields has the form

$$\hat{\boldsymbol{E}}^{(-)}(\boldsymbol{r},t) = \sum_{\mu} c_{\mu} \boldsymbol{e}_{\mu} e^{i\omega t - i\boldsymbol{k}_{\perp}\boldsymbol{r}_{\perp} - ik_z(\mu)z} \hat{a}_{\mu}^{+}(z). \tag{1}$$

Here, the sum is performed over all the field modes indicated by the index $\mu = \{\omega, \boldsymbol{k}_{\perp}, \alpha\}$, which includes the frequency, transverse components of the wave vector, and polarization index $\alpha$. $\hat{a}_{\mu}^{+}(z)$ is the photon creation operator in the corresponding field mode, $\boldsymbol{e}_{\mu}$ is the unit polarization vector, and $k_z(\mu) = \sqrt{\omega^2 n^2(\omega)/c^2 - k_\perp^2}$ is the longitudinal component of the wave vector. The expansion coefficients are given by the expression $c_\mu = \sqrt{\dfrac{2\pi\hbar\omega}{D^2 T c n(\omega)\cos\theta_\mu}}$, where $D$ is the transverse dimension of the quantization volume, and $\theta_\mu$ is the polar angle between the wave vector and the $z$ axis. In this case, the field operators satisfy the evolution equation

$$-i\hbar\frac{\partial \hat{a}_\mu^+}{\partial z} = \left[\hat{a}_\mu^+, \hat{G}_{nl}\right], \qquad (2)$$

where the nonlinear interaction operator $\hat{G}_{nl}(z)$ for PDC in the field of a classical pump wave $\boldsymbol{E}_p(\boldsymbol{r},t)$ at the frequency $\omega_p$ has the form

$$\hat{G}_{nl}(z) = 4\pi \int_{t_0}^{t_0+T} dt \iint d^2 r_\perp \chi^{(2)} \boldsymbol{E}_p(\boldsymbol{r},t) \hat{\boldsymbol{E}}_s^{(-)}(\boldsymbol{r},t) \hat{\boldsymbol{E}}_i^{(-)}(\boldsymbol{r},t) + h.c. \qquad (3)$$

Here, $\hat{\boldsymbol{E}}_s^{(-)}(\boldsymbol{r},t)$ and $\hat{\boldsymbol{E}}_i^{(-)}(\boldsymbol{r},t)$ are the fields of the signal and idler components of the scattered radiation with frequencies higher and lower than $\omega_p/2$, respectively, and $\chi^{(2)}$ is the tensor of the quadratic susceptibility of a nonlinear medium. Further, we assume that the pump is a Gaussian beam with diameter $d$. The pump field can be written as a superposition of plane waves, as follows:

$$\boldsymbol{E}_p(\boldsymbol{r},t) = E_0 \boldsymbol{e}_p e^{-r_\perp^2/d^2 + ik_{pz}z - i\omega_p t} = \frac{E_0 d^2}{4\pi} \boldsymbol{e}_p \int e^{-d^2 k_{p\perp}^2/4 + i\boldsymbol{k}_{p\perp}\boldsymbol{r}_\perp + ik_{pz}z - i\omega_p t} d^2 k_{p\perp}. \qquad (4)$$

If we substitute (1) and (4) into (3), perform an integration over time and transverse coordinates, and use the resulting delta functions, the expression for the nonlinear interaction operator is transformed into the quadratic form of the operators of creation and annihilation of photons in different modes of the scattered field.

$$\hat{G}_{nl}(z) = \hbar \frac{d^2}{D^2} \sum_{\omega_s, \boldsymbol{k}_{s\perp}, \boldsymbol{k}_{i\perp}} \chi_{eff}^{(2)} E_0 \frac{8\pi^3 \omega_s \omega_i}{c^2 \sqrt{k_s k_i \cos\theta_s \cos\theta_i}} e^{i\Delta k_z z - \frac{d^2(\boldsymbol{k}_{s\perp}+\boldsymbol{k}_{i\perp})^2}{4}} \hat{a}_s^+(z)\hat{a}_i^+(z) + h.c. \qquad (5)$$

Here, $\chi_{eff}^{(2)}$ is the result of convolution of the quadratic susceptibility tensor with unit polarization vectors, $\Delta k_z = k_{pz} - k_{sz} - k_{iz}$ is the longitudinal phase mismatch, and the idler mode frequency is uniquely determined by the energy conservation law $\omega_p = \omega_s + \omega_i$. The size of the quantization volume should disappear in the expressions for the observables when summation by modes is replaced by integration.

Scattered fields at different frequencies $\omega_s$ are independent. Therefore, the operator $\hat{G}_{nl}(z)$ can be considered at fixed frequencies of the signal and idler radiation. However, the angular modes are entangled; owing to the dependence of the exponents in (5) on combinations of angular variables, one idler mode is conjugated with a large number of signal modes, and vice versa. In principle, the quadratic form in (5) can be diagonalized; but owing to the dependence on the $z$ coordinate, the basis of the eigenmodes, in which this quadratic form is diagonal, differs at different points of the crystal. However, this study shows that under particular conditions, the eigenmodes of the operator in (5) are almost independent of the $z$ coordinate.

## III. ANGULAR DEPENDENCIES

Because we are interested in the angular structure of the biphoton field, we will retain only the angular dependencies and group all the other parameters into a single coefficient of parametric interaction $\gamma$. In addition, for convenience, we replace the summation over the transverse components of the wave vectors by an integration over the angles that determine the directions of the wave vectors.

$$\hat{G}_{nl}(z) = \hbar\gamma \iint \frac{\chi(\theta_s,\varphi_s,\theta_i,\varphi_i)}{\sqrt{\cos\theta_s \cos\theta_i}} e^{i\Delta k_z z - \frac{d^2(\mathbf{k}_{s\perp}+\mathbf{k}_{i\perp})^2}{4}} \hat{a}_s^+(z)\hat{a}_i^+(z) d\Omega_s d\Omega_i + h.c. \tag{6}$$

Here, $d\Omega_{s,i} = \sin\theta_{s,i} d\theta_{s,i} d\varphi_{s,i}$ are the elements of the solid angle, polar $\theta_{s,i}$ and azimuthal $\varphi_{s,i}$ angles determine the direction of the corresponding wave vectors $\mathbf{k}_{s,i}$ (Fig. 1), and the angular dependence of the effective quadratic susceptibility $\chi(\theta_s,\varphi_s,\theta_i,\varphi_i)$ is related to the anisotropy of the nonlinear crystal. We neglect the relatively weak angular dependence of the wave vector length.

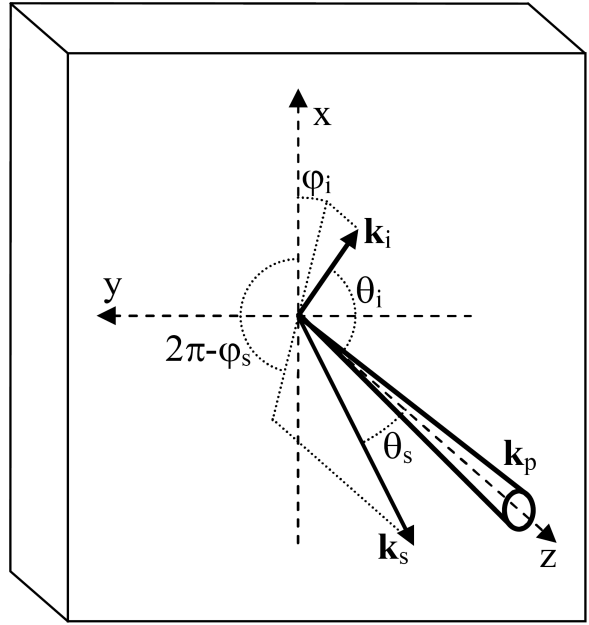

FIG. 1. Scattering geometry: the pump beam propagates along the axis z, the directions of the wave vectors of the signal $\mathbf{k}_s$ and idler $\mathbf{k}_i$ radiation are determined by the polar $\theta_{s,i}$ and azimuthal $\varphi_{s,i}$ angles.

Substituting (6) into (2), we obtain the evolution equations for the operators of the scattered field. The equations for the modes of the signal radiation have the form

$$\frac{\partial \hat{a}_s^+(z,\theta_s,\varphi_s)}{\partial z} = -i\gamma \iint \frac{\chi(\theta_s,\varphi_s,\theta_i,\varphi_i)}{\sqrt{\cos\theta_s \cos\theta_i}} e^{-i\Delta k_z z - \frac{d^2(\mathbf{k}_{s\perp}+\mathbf{k}_{i\perp})^2}{4}} \hat{a}_i(z,\theta_i,\varphi_i) d\Omega_i, \tag{7}$$

and the equations for the idler radiation are similar. In the case of low parametric gain $\gamma L \ll 1$, where $L$ is the length of the crystal, we can immediately obtain a solution for the SPDC, as follows:

$$\hat{a}_s^+(L,\theta_s,\varphi_s) = \hat{a}_s^+(0,\theta_s,\varphi_s) + \iint F^*(\theta_s,\varphi_s,\theta_i,\varphi_i) \hat{a}_i(0,\theta_i,\varphi_i) d\Omega_i, \tag{8}$$

where the function

$$F(\theta_s, \varphi_s, \theta_i, \varphi_i) = i\gamma L \frac{\chi(\theta_s, \varphi_s, \theta_i, \varphi_i)}{\sqrt{\cos\theta_s \cos\theta_i}} e^{\frac{i\Delta k_z L}{2} - \frac{d^2(\mathbf{k}_{s\perp} + \mathbf{k}_{i\perp})^2}{4}} \operatorname{sinc}\left(\frac{\Delta k_z L}{2}\right) \quad (9)$$

is typically called the biphoton amplitude. It can be used to express the wave function of the state of the scattered field at the exit of the crystal in the case of SPDC [27], as follows:

$$|\psi(L)\rangle = \left\{ \iint F(\theta_s, \varphi_s, \theta_i, \varphi_i) \hat{a}_s^+(\theta_s, \varphi_s) \hat{a}_i^+(\theta_i, \varphi_i) d\Omega_s d\Omega_i \right\} |vac\rangle, \quad (10)$$

As is known, a two-photon state of this type can be decomposed into Schmidt modes

$$|\psi(L)\rangle = \left\{ \sum_j \sqrt{\lambda_j} \hat{b}_j^+ \hat{c}_j^+ \right\} |vac\rangle, \quad (11)$$

where

$$\begin{aligned}
\hat{b}_j^+ &= \int v_j(\theta_i, \varphi_i) \hat{a}_i^+(\theta_i, \varphi_i) d\Omega_i \\
\hat{c}_j^+ &= \int \tilde{v}_j(\theta_s, \varphi_s) \hat{a}_s^+(\theta_s, \varphi_s) d\Omega_s
\end{aligned} \quad (12)$$

are the operators of photon creation in the Schmidt modes. If the angular dependence of $\frac{\chi(\theta_s, \varphi_s, \theta_i, \varphi_i)}{\sqrt{\cos\theta_s \cos\theta_i}}$ is neglected in the expression for the biphoton amplitude (9), and the sinc function is approximated by a Gaussian function, then the decomposition of the function in (10) into Schmidt modes can be performed analytically [22,23]. This approximation is acceptable if the frequencies of the signal and idler radiation are comparable and the angular width of the phase matching is sufficiently small. However, in the case of strongly non-degenerate PDC, when the frequency of the idler radiation is in the terahertz range and significantly less than that of the signal radiation, the angular width of the phase matching can be of the order of units or even tens of degrees [18]. In this case, the study of the mode structure of scattered radiation requires the consideration of all angular dependencies.

Let us return to the expression for the nonlinear interaction operator in (6) at an arbitrary parametric gain. The angular dependencies of the combinations of the wave vectors are as follows:

$$(\mathbf{k}_{s\perp} + \mathbf{k}_{i\perp})^2 = \underbrace{(k_i \sin\theta_i - k_s \sin\theta_s)^2}_{\Delta k_\perp^2} + 4 k_i k_s \sin\theta_i \sin\theta_s \cos^2\left(\frac{\varphi_i - \varphi_s}{2}\right) \quad (13)$$

and

$$\Delta k_z = k_p \cos\theta_p - (k_i \cos\theta_i + k_s \cos\theta_s). \quad (14)$$

If we consider the smallness of the angles $\theta_p$ and the exact phase-matching condition in the transverse direction $\mathbf{k}_{p\perp} = \mathbf{k}_{s\perp} + \mathbf{k}_{i\perp}$, we can approximate

$$\Delta k_z \approx \underbrace{k_p - (k_i \cos\theta_i + k_s \cos\theta_s)}_{\Delta k_{z0}} - \frac{(\mathbf{k}_{s\perp} + \mathbf{k}_{i\perp})^2}{2 k_p}. \quad (14a)$$

Then, the exponent in expression (6) has the form

$$i\Delta k_z z - \frac{d^2(\mathbf{k}_{s\perp} + \mathbf{k}_{i\perp})^2}{4} = i\left(\Delta k_{z0} - \frac{\Delta k_\perp^2}{2 k_p}\right) z - \frac{d^2 \Delta k_\perp^2}{4} - 2\tau\left(1 + i\frac{2z}{d^2 k_p}\right) \cos^2\left(\frac{\varphi_i - \varphi_s}{2}\right), \quad (15)$$

where $\tau = d^2 k_i k_s \sin\theta_i \sin\theta_s / 2$. The scattering maximum is observed in the directions determined by the complete phase-matching condition at which all three terms in this expression are equal to zero. The first two terms determine the polar angles $\theta_{s,i}$, and the last requires that the azimuthal angles $\varphi_{s,i}$ differ from each other by $\pi$.

For numerical estimates, as in [18], we consider PDC in a lithium niobate crystal with a thickness of $L = 1$ mm under the action of a pump beam with a wavelength of $\lambda_p = 523.3$ nm and diameter of $d = 300$ μ$m$. In lithium niobate, the polar angle of the idler wave in the direction of exact phase matching is approximately equal to $60°$ until its frequency approaches the resonance at the frequency of the lower optical phonon of 7.44 $THz$ [9]. Near the phase-matching condition, the parameter $\tau \approx (k_i d \sin\theta_i)^2 / 2$ changes from $\sim 0.04$ to $\sim 1600$ when the idler wave frequency changes from 0.01 to 2 THz. For the imaginary part in the last term in (15), the inequalities $2z/d^2 k_p \ll 1$ and $4\tau z/d^2 k_p \leq 4\tau L/d^2 k_p < \pi$ are satisfied. Thus, in the last term, the imaginary part is always much less than the real one, and as long as it does not lead to oscillations, it can be neglected (the approximation becomes incorrect only near a frequency of 2 THz). Under other experimental conditions, the values of the parameter $\tau$ may differ. However, since $\tau$ decreases with decreasing frequency of the idler radiation, then for a strongly non-degenerate PDC there always exists a range of sufficiently low idler radiation frequencies for which our approximations are correct. Then, the operator of the nonlinear interaction in (6) can be rewritten as

$$\hat{G}_{nl} \approx \hbar\gamma \int \left\{ e^{i(\Delta k_{z0} - \Delta k_\perp^2/2k_p)z - d^2\Delta k_\perp^2/4} \frac{\sin\theta_s \sin\theta_i}{\sqrt{\cos\theta_s \cos\theta_i}} d\theta_s d\theta_i \times \right.$$
$$\left. \times \left[ \int_0^{2\pi} \chi(\theta_s, \varphi_s, \theta_i, \varphi_i) e^{-\tau\{1+\cos(\varphi_i - \varphi_s)\}} \hat{a}_i^+ \hat{a}_s^+ d\varphi_s d\varphi_i \right] + h.c. \right\} \quad (16)$$

The part of expression (16) in square brackets contains all dependencies on the azimuthal angles and does not explicitly depend on $z$. Therefore, we can diagonalize the nonlinear interaction operator in the azimuthal angle space. Clearly, there will remain entanglement in the polar angles, depending on the coordinate $z$. However, it was shown in [18] that in the case of a strongly non-degenerate PDC, when the idler radiation frequency is in the terahertz range, the Fedorov parameter for the polar modes, which provides an estimation of the number of Schmidt modes, is very close to unity. This indicates that there is practically no entanglement in the polar angles. The dependence on the polar angles is reduced to the shape of a single pair of polar modes of the signal and idler radiation. The maximum of these polar modes is concentrated near the angles corresponding to the exact phase-matching condition. Therefore, we can restrict ourselves to considering the operator in (16) in the case of exact phase matching for polar angles $\Delta k_{z0} = \Delta k_\perp = 0$ and focus on the structure of the azimuthal modes.

In [18], the angular dependence of the effective quadratic susceptibility was used in the form
$$\chi_1(\theta_s, \varphi_s, \theta_i, \varphi_i) = \cos\varphi_s \cos\varphi_i. \quad (17)$$

A more accurate expression for the convolution of the quadratic susceptibility tensor with the polarization vectors of the fields in the eee-geometry PDC has the form

$$\chi(\theta_s, \varphi_s, \theta_i, \varphi_i) = \cos\theta_s \cos\theta_i \sqrt{\left(1 + \text{tg}^2\theta_s \cos^2\varphi_s\right)\left(1 + \text{tg}^2\theta_i \cos^2\varphi_i\right)}. \tag{18}$$

If we consider that $\theta_s \ll 1$ and $\theta_i \approx 60°$, we obtain $\chi(\theta_s, \varphi_s, \theta_i, \varphi_i) \approx \frac{1}{2}\sqrt{1 + 3\cos^2\varphi_i}$. This dependence is well approximated by the expression

$$\chi_2(\theta_s, \varphi_s, \theta_i, \varphi_i) \approx \frac{1}{2}\left(1 + \cos^2\varphi_i\right). \tag{19}$$

This approximation will significantly simplify the procedure for obtaining eigenmodes. Further, we consider the diagonalization of the operator in (16) in azimuthal angles for dependencies $\chi_1$ and $\chi_2$.

## IV. AZIMUTAL EIGENMODES AND SCATTERING MATRIX

Let us decompose the field operators into a Fourier series in azimuthal angles, as follows: $\hat{a}^+_{s,i}(\varphi_{s,i}) = \frac{1}{\sqrt{2\pi}} \sum_n \hat{a}_n^{+(s,i)} e^{in\varphi_{s,i}}$, where $\hat{a}_n^{+(s,i)} = \frac{1}{\sqrt{2\pi}} \int_{-\pi}^{\pi} \hat{a}^+_{s,i}(\varphi_{s,i}) e^{-in\varphi_{s,i}} d\varphi_{s,i}$ are the operators of photon creation in the Fourier modes. Substituting this expansion into (16), we obtain

$$\hat{G}_{nl} \sim \int_0^{2\pi} \chi_\alpha(\varphi_s, \varphi_i) e^{-\tau\{1+\cos(\varphi_i - \varphi_s)\}} \hat{a}_i^+(\varphi_i) \hat{a}_s^+(\varphi_s) d\varphi_s d\varphi_i + h.c.$$
$$= \frac{e^{-\tau}}{2\pi} \sum_{n,m} \hat{a}_n^{+(i)} \hat{a}_m^{+(s)} \int_0^{2\pi} \chi_\alpha(\varphi_s, \varphi_i) e^{in\varphi_i + im\varphi_s - \tau\cos(\varphi_i - \varphi_s)} d\varphi_s d\varphi_i + h.c. \tag{20}$$

It can be observed that it is convenient to make a substitution using the difference $\psi = \varphi_i - \varphi_s$ and the half-sum $\Phi = (\varphi_i + \varphi_s)/2$ of the angles. The functions $\chi_\alpha(\varphi_s, \varphi_i)$ in both cases can also be represented as combinations of exponents from angles $\psi$ and $\Phi$:

$$\hat{G}_{nl} \sim \frac{e^{-\tau}}{2\pi} \sum_{n,m} \hat{a}_n^{+(i)} \hat{a}_m^{+(s)} \int_0^{2\pi} \chi_\alpha\left(\Phi - \frac{\psi}{2}, \Phi + \frac{\psi}{2}\right) e^{i(n+m)\Phi + i(n-m)\psi/2 - \tau\cos\psi} d\psi d\Phi + h.c. \tag{21}$$

By integrating over angle $\Phi$, we obtain the relationships between $n$ and $m$, and by integrating over angle $\psi$, we obtain the Infeld functions $I_k(\tau) = \frac{1}{\pi}\int_0^\pi e^{\tau\cos\psi} \cos(k\psi) d\psi$ of various orders. As a result, the nonlinear interaction operator in the Fourier-mode basis has the form

$$\hat{G}_{nl} = \hbar\tilde{\gamma} \sum_{n,m} H_{nm} \hat{a}_n^{+(i)} \hat{a}_{-m}^{+(s)} + h.c., \tag{22}$$

where the matrix $\{H_{nm}\}$ for two different angular dependencies of the quadratic susceptibility $\chi_\alpha(\varphi_s, \varphi_i)$ is as follows:

$$H_{nm}^{(1)} = \frac{(-1)^n}{2} e^{-\tau} \left\{\left(I_{n+1}(\tau) + I_{n-1}(\tau)\right)\delta_{n,m} + I_{n+1}(\tau)\delta_{n+2,m} + I_{n-1}(\tau)\delta_{n-2,m}\right\}, \tag{23}$$

for the first case in (17), and for the second case in (19),

$$H_{nm}^{(2)} = (-1)^n e^{-\tau} I_m(\tau) \left\{\delta_{n,m} + \frac{1}{6}\delta_{n+2,m} + \frac{1}{6}\delta_{n-2,m}\right\}. \tag{24}$$

In both the cases, this matrix consists of two tridiagonal parts, for even and odd values of $n$.

Thus, instead of integrating over the continuous values of the azimuthal angles, we obtain the summation over the discrete index $n$. Elements of matrices $H_{nm}$ are expressed in terms of the values $e^{-\tau}I_n(\tau)$. For sufficiently large $n$, they tend to zero. Therefore, for any value of the parameter $\tau$, it is sufficient to analyze a finite-size matrix $|n|,|m| \leq n_{\max}$ instead of an infinite one. After that, the quadratic form (22) is easy to diagonalize numerically by representing the matrix as:

$$H_{nm} = \sum_j W_{nj}^{-1} R_j V_{jm}, \tag{25}$$

where $\{V_{jm}\}$ and $\{W_{jm}\}$ are the orthogonal matrices (for the case of a symmetric matrix $H_{nm}^{(1)}$ $V = W$), and $R_j$ are the eigenvalues of the original matrix $\{H_{nm}\}$. As a result, the nonlinear interaction operator has the form

$$G_{nl} = \hbar\tilde{\gamma}\sum_j R_j \left(\sum_n W_{jn}\hat{a}_n^{+(i)}\right)\left(\sum_m V_{jm}\hat{a}_{-m}^{+(s)}\right) + h.c. \tag{26}$$

Here, $\hat{b}_j^+ = \sum_n W_{jn}\hat{a}_n^{+(i)}$ and $\hat{c}_j^+ = \sum_m V_{jm}\hat{a}_{-m}^{+(s)}$ are the operators of photon creation in the eigenmodes of the idler and signal radiation, respectively, and the squared eigenvalues $R_j^2$ correspond to the parameters $\lambda_j$ in the Schmidt decomposition (11). The solution of their evolution equations has the form of the Bogolyubov transformations:

$$\begin{aligned}\hat{b}_j^+(L) &= \hat{b}_j^+(0)\cosh g_j - i\hat{c}_j(0)\sinh g_j \\ \hat{c}_j^+(L) &= \hat{c}_j^+(0)\cosh g_j - i\hat{b}_j(0)\sinh g_j,\end{aligned} \tag{27}$$

where $g_j = \tilde{\gamma}LR_j$. Thus, we obtain the basis of the eigenmodes. In this basis, the nonlinear interaction operator and the scattering matrix have a diagonal form; i.e., we performed the Bloch-Messiah decomposition [24]. Because the shape of the azimuthal eigenmodes is not dependent on the parametric gain, these modes coincide with the Schmidt modes in (12) for the case of SPDC at $\tilde{\gamma}L \ll 1$.

Using inverse transformations, one can express the scattering matrix in the original basis of the plane waves in terms of the elements of the unitary matrices $U_j(\varphi_i) = (2\pi)^{-1/2}\sum_n e^{in\varphi_i}W_{jn}$, $\tilde{U}_j(\varphi_s) = (2\pi)^{-1/2}\sum_m e^{im\varphi_s}V_{jm}$ :

$$\begin{aligned}\hat{a}_i^+(\varphi_i, L) &= \sum_j U_j(\varphi_i)\left\{\hat{b}_j^+(0)\cosh g_j - i\hat{c}_j(0)\sinh g_j\right\} \\ \hat{a}_s^+(\varphi_s, L) &= \sum_j \tilde{U}_j^*(\varphi_s)\left\{\hat{c}_j^+(0)\cosh g_j - i\hat{b}_j(0)\sinh g_j\right\}\end{aligned} \tag{28}$$

In the case of other experimental conditions and angular dependencies of the quadratic susceptibility, the matrix $\{H_{nm}\}$ may differ from our expressions (23,24). However, our method to obtain the scattering matrix is applicable in the general case of strongly non-degenerate PDC for arbitrary values of the parametric gain.

## V. SHAPE AND EFFECTIVE NUMBER OF EIGENMODES

Any quantum statistical properties of the scattered field can be determined using the scattering matrix. In particular, the azimuthal dependence of the intensity of the idler scattered radiation has the form

$$I_i(\varphi_i) \sim \langle vac | \hat{a}_i^+(\varphi_i, L) \hat{a}_i(\varphi_i, L) | vac \rangle = \sum_j \sinh^2 g_j |U_j(\varphi_i)|^2, \qquad (29)$$

i.e. it consists of the sum of intensities of the eigenmodes $|U_j(\varphi_{i,s})|^2$ with the weights $\sinh^2 g_j$. A similar expression can be written for the signal radiation intensity. The effective number of modes can be determined as follows:

$$K = \left(\sum_j \sinh^2 g_j\right)^2 \Big/ \sum_j \sinh^4 g_j. \qquad (30)$$

In the case of SPDC at a small parametric gain $\tilde{\gamma} L \ll 1$, the weights are proportional to the squared eigenvalues $R_j^2$, and the parameter $K$ is equal to the Schmidt number [20].

Figures 2 and 3 show the calculated angular dependencies of the intensities of the azimuthal eigenmodes of the idler radiation at different terahertz frequencies for two different angular dependencies of the effective quadratic susceptibility $\chi_\alpha(\varphi_s, \varphi_i)$. The shape of the eigenmodes of the signal radiation in case $\chi_1$ is exactly the same, and in case $\chi_2$ it is slightly different, but this difference is noticeable only for sufficiently low frequencies $f_i \leq 0.05$ $THz$. For clarity, the intensity of each eigenmode in the graphs is shifted by its eigenvalue $R_j$: $I_j(\varphi_i) = R_j \left(1 + 2\pi |R_j| |U_j(\varphi_i)|^2\right)$. Positive and negative eigenvalues correspond to the even and odd parts of the matrices in (23) and (24).

At low frequency $f_i = 0.01$ $THz$, the parameter $\tau \ll 1$, and $I_n(\tau) \ll I_0(\tau) \approx 1$. Therefore, in the matrices in (23) and (24), almost all elements are approximately zero, and all the scattered radiation is concentrated in a single azimuthal mode (Figs 2a and 3a). As the frequency $f_i$ and parameter $\tau$ increase, the difference between the Infeld functions of different orders decreases, and the weights of the other modes grow (Figs 2b and 3b). Therefore, the difference between the even and odd parts of the matrices gradually disappears. At frequency $f_i = 0.1$ $THz$, the modes corresponding to the positive and negative eigenvalues are similar (Figs 2c and 3c). The two-fold degeneracy of the system is explained by the fact that the matrix consists of two parts — for even and odd terms — and for $\tau \gg 1$, they are approximately the same. With a further increase in the idler radiation frequency, the widths of the modes decrease, and their numbers increase (Figs 2d and 3d).

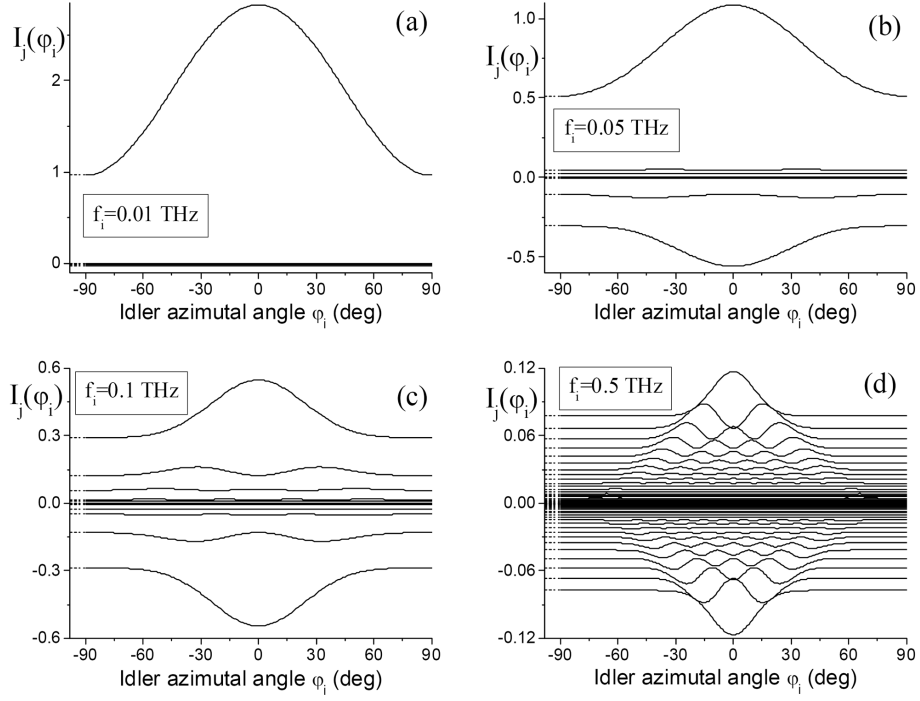

FIG. 2. The shape of the azimuthal eigenmodes of the idler radiation at different values of the terahertz frequency $f_i$ in the case of the angular dependence of the effective quadratic susceptibility $\chi_1(\varphi_i, \varphi_s)$ (17). For clarity, the intensity of each eigenmode is shifted by its eigenvalue $R_j$: $I_j(\varphi_i) = R_j\left(1 + 2\pi|R_j|\,|U_j(\varphi_i)|^2\right)$. The eigenvalue level is shown by the dotted line on the left.

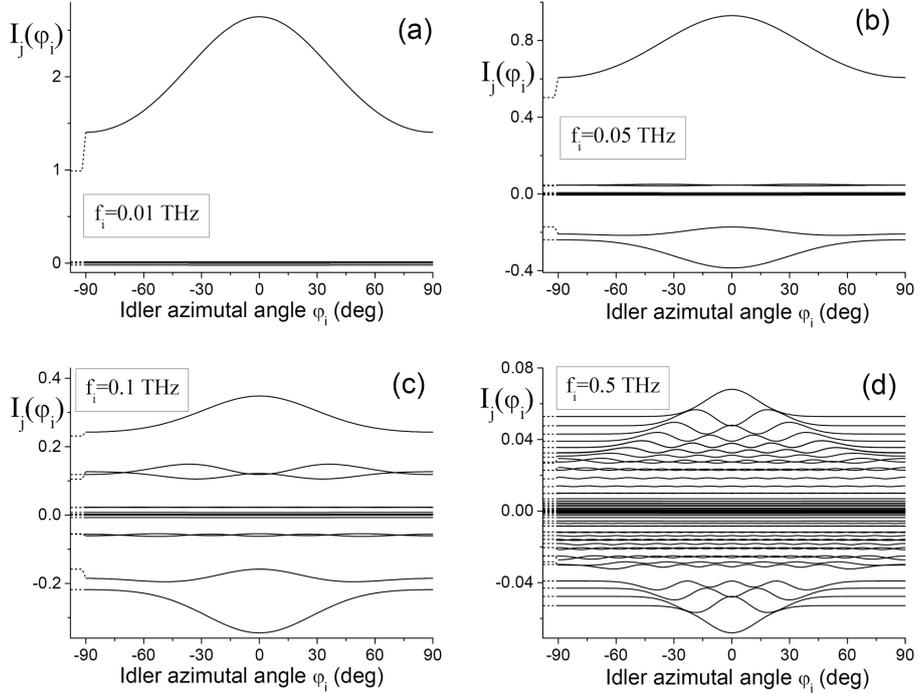

FIG. 3. The shape of the azimuthal eigenmodes of the idler radiation at different values of the terahertz frequency $f_i$ in the case of the angular dependence of the effective quadratic susceptibility $\chi_2(\varphi_i, \varphi_s)$ (19). For clarity, the intensity of each eigenmode is shifted by its eigenvalue $R_j$: $I_j(\varphi_i) = R_j\left(1 + 2\pi|R_j|\,|U_j(\varphi_i)|^2\right)$. The eigenvalue level is shown by the dotted line on the left.

Figure 4 shows the dependencies of the effective number of modes $K$ on the idler radiation frequency at various parametric gains. In the case $\tilde{\gamma}L \ll 1$, the number of modes increases linearly with frequency, and at frequency $f_i = 2\ THz$, it is already of the order of several tens. Note that the azimuthal Schmidt number is approximately twice the Fedorov parameter obtained in [18], which is owing to the already mentioned two-fold degeneracy.

In the case of stimulated PDC, when the parametric gain for separate modes is $g_j \geq 1$, the effective number of modes decreases because the weights of the modes $\sinh^2 g_j$ with the large eigenvalues increase at a faster rate. At a constant pump intensity, the highest values $g_{max} = \tilde{\gamma}LR_{max}$ are practically independent of frequency. This is explained by the fact that although the largest eigenvalues $R_{max}$ decrease with increasing frequency, the coefficient $\tilde{\gamma}$ is proportional to the idler radiation frequency (5). As a result, in Fig. 4, the effective number of modes in the case of stimulated PDC grows more slowly than in the case of SPDC.

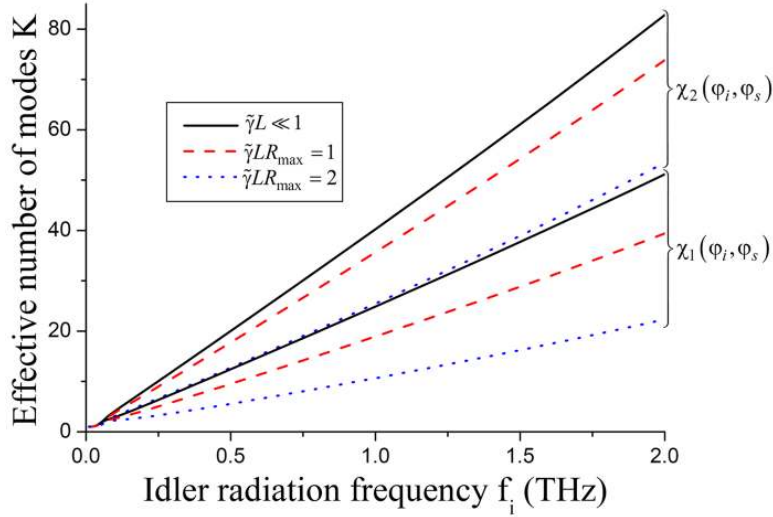

FIG. 4. Dependencies of the effective number of modes $K$ (30) on the idler radiation frequency $f_i$ for different parametric gains $\tilde{\gamma}L$ in the cases of two different angular dependencies of the effective quadratic susceptibility $\chi_\alpha(\varphi_i, \varphi_s)$.

## VI. CONCLUSION

Summarizing, we obtained azimuthal eigenmodes for the nonlinear interaction operator in (6), which describes strongly non-degenerate PDC. These eigenmodes allowed us to obtain the scattering matrix in (28), which describes the quantum state of the scattered radiation for an arbitrary parametric gain.

The angular structure and the relative weights of the azimuthal eigenmodes allow to optimize the measurement of the quantum correlation properties of the optical-terahertz biphoton field. The excess over unity of the normalized correlation function of the signal and idler radiation intensities $g_2$ is inversely proportional to the number of detected modes [18]. The angular dependences in Figures 2 and 3 show that single-mode optical-terahertz biphotons can be obtained only at very low frequencies of terahertz idler radiation. At higher terahertz frequencies, detectors will always register multimode radiation. However, the choice of the angular aperture of the detectors equal to the width of the mode with the largest weight can significantly reduce the effective number of detected modes and, thus, increase the level of measured correlation.

At the same time, the effect of two-mode squeezing, which consists of the suppressing of fluctuations in the photon-number difference of the idler and signal radiation below the classical limit, does not depend on the number of detected modes [28]. The value of the two-mode squeezing is sensitive to the matching of the detected modes of the idler and signal radiation. According to the structure of azimuthal modes obtained by us, to measure the effect of two-mode squeezing in an optical-terahertz biphoton field, the angular apertures of the signal and idler radiation detectors in the azimuthal angles should be equal and sufficiently large. Under this condition, the detectors will collect a large number of conjugate modes. At the same time, the angular apertures of the detectors in the polar angles should differ significantly in accordance with the difference in the angular widths of the polar modes of the optical signal and terahertz idler radiation [18].

The possibility of measuring both the quantum correlations between optical and terahertz photons in the SPDC, and the two-mode squeezing in an optical-terahertz biphoton field at large values of the parametric gain can be used to implement quantum-optical technologies in the terahertz range, in particular, to create methods of calibrating the sensitivity and quantum efficiency of terahertz detectors [16,29,30].

## ACKNOWLEDGMENTS

The authors thank G. Kh. Kitaeva for fruitful discussions. This work was performed under financial support of the Russian Science Foundation (Grant No. 17-12-01134).

This is a preprint of an article published in Applied Physics B. The final authenticated version is available online at: https://doi.org/10.1007/s00340-021-07634-5.